\documentclass[10pt,twocolumn,letterpaper]{article}

\usepackage{cvpr}              % To produce the CAMERA-READY version
\definecolor{cvprblue}{rgb}{0.21,0.49,0.74}
\usepackage[pagebackref,breaklinks,colorlinks,allcolors=cvprblue]{hyperref}

\usepackage[linesnumbered,ruled,vlined]{algorithm2e}

\def\paperID{*****} % *** Enter the Paper ID here
\def\confName{CVPR}

\title{Bridging the Gap between 

Micro-scale Traffic Simulation and 4D Digital Cityscapes}

\author{Longxiang Jiao*\\
ETH Zurich\\
{\tt\small ljiao01@ethz.ch}
\and
Lukas Hofmann*\\
ETH Zurich\\
{\tt\small luhofmann@ethz.ch}
\and
Yiru Yang*\\
University of Zurich\\
{\tt\small yiru.yang@uzh.ch}
\and
Zhanyi Wu*\\
University of Zurich\\
{\tt\small zhanyi.wu@uzh.ch}
\and
Jonas Egeler*\\
ETH Zurich\\
{\tt\small jegeler@ethz.ch}
}

\toggletrue{cvprfinal}

\begin{document}
\maketitle

% \textbf{Content}
% \begin{itemize}
%     \item Abstract
%     \item Introduction
%     \item Related work
%     \item TraCI bridge
%     \item OSC/JUCE auralization
%     \item Traffic lights
%     \item Optimizations
%     \item User study
%     \item Conclusion and Future work
%     \item References
%     \item Somewhere geneva stuff or smth? If there's space
% \end{itemize}
% \clearpage

\begin{abstract}
While micro-scale traffic simulations provide essential data for urban planning, they are rarely coupled with the high-fidelity visualization or auralization necessary for effective stakeholder communication. In this work, we present a real-time 4D visualization framework that couples the SUMO traffic with a photorealistic, geospatially accurate VR representation of Zurich in Unreal Engine 5. Our architecture implements a robust C++ data pipeline for synchronized vehicle visualization and features an Open Sound Control (OSC) interface to support external auralization engines. We validate the framework through a user study assessing the correlation between simulated traffic dynamics and human perception. Results demonstrate a high degree of perceptual alignment, where users correctly interpret safety risks from the 4D simulation. Furthermore, our findings indicate that the inclusion of spatialized audio alters the user's sense of safety, showing the importance of multimodality in traffic simulations.
\end{abstract}    
\section{Introduction}
\label{sec:intro}

Urban planning and transportation engineering benefit significantly from simulation tools which predict the impact of infrastructural changes. As cities become denser, the ability to model vehicle flow is critical for sustainable development. Microscale traffic simulators, such as Simulation of Urban Mobility (SUMO) \cite{SUMO2018}, have become the standard for these tasks, offering a robust way to model individual vehicle behaviors.

\begin{figure}
    \centering
    \includegraphics[width=0.75\linewidth]{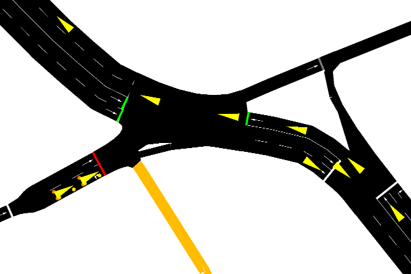}
    \caption{The default 2D visualization interface provided by SUMO. The traffic flow is represented only as moving shapes on a 2D map.}
    \label{fig:2d}
\end{figure}

However, standard tools currently lack the power to visualize and auralize such traffic simulation in an immersive manner. This creates a barrier for urban planners when communicating their findings to stakeholders, policy makers, or the general public. While frameworks like SUMO excel at generating logical data, their native visualization capabilities are typically abstract and limited to moving dots on a 2D map. \cref{fig:2d} shows the SUMO's default visualization GUI. 

\begin{figure}
    \centering
    \includegraphics[width=0.75\linewidth]{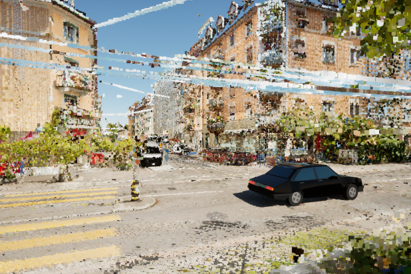}
    \caption{Our proposed framework in Unreal Engine 5, visualizing the same traffic simulation as 3D dynamic vehicles within a geospatial reconstruction of Zurich.}
    \label{fig:3d}
\end{figure}

To address this limitation, we present a visualization framework that bridges the gap between abstract traffic logic and immersive rendering in digital cityscapes. By coupling SUMO with Unreal Engine 5 (UE5) \cite{unrealengine}, we create a 3D visualization of traffic in the city of Zurich which is explorable in Virtual Reality (VR). Our system utilizes the Traffic Control Interface (TraCI) protocol \cite{TraCI} to stream simulation data from a SUMO server in real-time, populating a geospatial reconstruction of the city with dynamic vehicles. A screenshot of the result is shown in \cref{fig:3d}.

Furthermore, we extend the simulation scope to include auditory feedback, as urban environments are defined not just by how they look, but also how they sound. Our framework includes an interface for Open Sound Control (OSC) \cite{OSC}, allowing the system to stream vehicle data to external audio engines for auralization \cite{Vorländer2020Acoutics} of traffic sounds. 

Finally, we demonstrate the effectiveness of our framework through a user study. We investigate the general immersion provided by the VR representation and the specific impact of audio on the user's perception of safety. Additionally, we evaluate the system's ability to accurately convey changes in traffic dynamics to the user.

Our main contributions are: 1) A C++ implementation linking SUMO’s traffic logic with Unreal Engine 5, applied to a VR geospatial representation of Zurich; 2) Support for integrating external auralization engines via OSC; 3) User study evaluating the environmental immersion, the influence of audio, and the perceptual alignment between user experience and simulated traffic dynamics.    

\section{Related Work}
\label{sec:related_work}

\subsection{Digital Twins for Urban Planning} 
Historically, urban planning has relied on 2D Geographic Information Systems (GIS) and static 3D renderings to communicate design intent. However, these tools often fail to convey the experiential qualities of a city, such as scale and atmosphere, to stakeholders. Recently, the adoption of Game Engines (e.g., Unreal Engine, Unity) has enabled the creation of ``Digital Twins" that integrate geospatial data with realistic rendering \cite{ijgi12080310}. Studies such as Dembski et al. \cite{su12062307} have demonstrated that immersive VR environments significantly improve public participation by allowing citizens to walk through future scenarios, rather than merely viewing them from a top-down perspective.

\subsection{Microscale Traffic Simulation}
Microscale traffic simulation refers to the modeling of traffic systems at the level of individual vehicles and their behaviour. One of the most widely adopted open-source microscale traffic simulators is Simulation of Urban Mobility (SUMO) \cite{SUMO2018}, which advances the simulation in fixed discrete time steps, numerically updating vehicle states at each step. 

SUMO can be run as a TCP server, which listens for commands to advance its simulation, instead of doing so on its own. At every timestep, the client can make additional queries requesting detailed traffic data, such as the positions of vehicles or the state of traffic lights. TCP packets exchanged with the SUMO server follow the Traffic Control Interface (TraCI) \cite{TraCI}, a client-server communication protocol designed for real-time traffic simulation control and information exchange.

While SUMO provides robust analytical data, its native visualization tools are abstract and limited to 2D. To bridge this gap, Olaverri-Monreal et al. \cite{s18124399} and Pechinger et al. \cite{Pechinger_Lindner_2024} have introduced interfaces connecting SUMO with the Unity game engine. Our work follows a similar paradigm, but leverages Unreal Engine 5 for higher visual realism and includes the additional auralization aspect.

\subsection{Auralization} 
A visual representation alone is insufficient for urban planning, as the acoustic environment plays a critical role in city liveability. Auralization \cite{Vorländer2020Acoutics, Auralization2015, unity_integrating_sound} extends visualization by simulating sound propagation based on physical data. A simple way to decouple an auralization implementation from the rendering engine is message passing via OSC \cite{OSC}, a protocol widely used to transmit time-critical control data over the network between audio devices.

\section{Methodology}
\label{sec:method}
\begin{figure}
    \centering
    \includegraphics[width=1\linewidth]{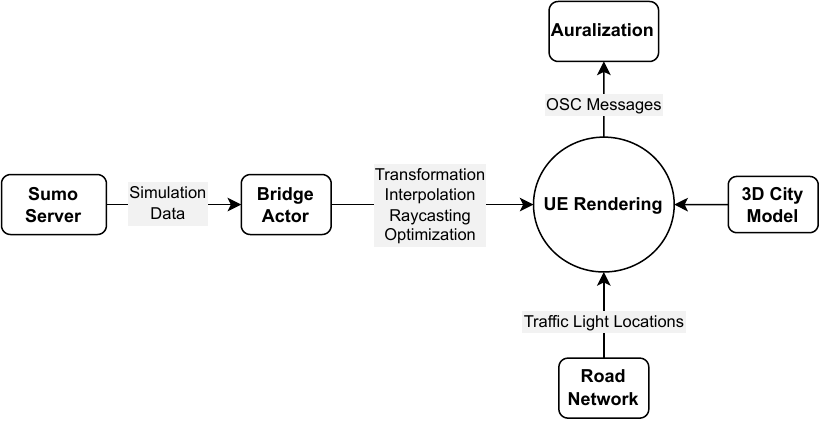}
    \caption{Architecture overview. The Bridge Actor ingests raw traffic data from the SUMO Server, applies geospatial transformations and interpolation for smooth visualization in Unreal Engine, and streams vehicle states via OSC for external auralization.} 
    \label{fig:system_overview}
\end{figure}

We present a real-time framework designed to synchronize SUMO traffic data with a photorealistic VR environment in UE5. As illustrated in Figure \ref{fig:system_overview}, most of the processing load falls on the Bridge Actor, which serves as a data mediator between the SUMO Server and the UE5 Renderer. The system operates by ingesting raw simulation data via a TCP socket in the form of TraCI packets, which are then processed through various transformations involving coordinate conversion, vehicle position interpolation, and raycast-based height finding to dynamically align the vehicles with the 3D City Model. Beyond visualization, vehicle states are streamed via OSC messages to an external JUCE Auralization engine, which simulates traffic sounds. Our code will be published by the research group at a later time. % TODO: Clean up git repo

\subsection{Asynchronous Producer-Consumer Approach}
\label{sec:prodcons}
A critical challenge in coupling traffic simulation with game engines is handling the blocking nature of network communication without freezing the rendering thread. To address this, we take an asynchronous producer-consumer approach in the Bridge Actor, where all network operations are performed by a dedicated producer thread, while processing and rendering is handled by the consumer thread.

At the start of the simulation, the producer thread establishes a TCP connection between UE5 and the SUMO server. It then requests the SUMO server to advance the simulation steps as well as sends queries to retrieve current traffic data at a fixed $N$~Hz rate. Correspondingly, the server uses a timestep of $\frac{1}{N}$ seconds, ensuring the original timescale of the simulation is preserved. $N$ is chosen to be low enough such that the server has enough time relative to its computational power to keep up with the amount of requests, but high enough to ensure a single timestep is not too large, which may degrade the quality of the simulation. The retrieved traffic data include newly departed vehicles, newly arrived vehicles, the positions of the vehicles, as well as the state of traffic lights. At each timestep, received new data is sent into a thread-safe map between vehicle IDs and vehicles states, shared with the consumer thread.    

Since the consumer thread handles rendering, it runs at a variable update frequency corresponding to the framerate from the game engine. The shared map approach decouples the two threads, preventing arbitrary speedups or slowdowns in the simulation tied to framerate fluctuations. 

\subsection{Vehicle Data Processing}
\subsubsection{Interpolation}
Conceptually, to ensure a timescale consistent with the SUMO simulation, it is sufficient for the consumer thread to simply always read the latest available update for each vehicle and traffic light from the shared map, and discard the others. However, this would result in choppy and unnatural vehicle movement. To prevent this, we implement linear interpolation. The system stores a PreviousState and TargetState for each vehicle. During a render cycle, the visual position is interpolated between the two states based on the elapsed time, producing smooth continuous motion.

\begin{algorithm}[t]
\caption{Vehicle State Interpolation}
\label{alg:interpolation}
% Define Function keyword style
\SetKwProg{Fn}{Function}{}{}
% Global variables initialization
$T_{\text{step}} \leftarrow$ SUMO step length\;
$V \leftarrow$ Map$\langle$VehicleID, VehicleState$\rangle$\;
\BlankLine
\Fn{\textsc{OnNetworkUpdate}($Data_{\text{new}}$)}{
    \ForEach{$v \in Data_{\text{new}}$}{
        $V[v].\text{Previous} \leftarrow V[v].\text{Target}$\;
        $V[v].\text{Target} \leftarrow \textsc{Parse}(v)$\;
        $V[v].\alpha \leftarrow 0.0$\;
    }
}
\BlankLine
\Fn{\textsc{TickVisuals}($\Delta t$)}{
    \ForEach{$v \in V$}{
        $v.\alpha \leftarrow v.\alpha + \frac{\Delta t}{T_{\text{step}}}$\;
        $v.\alpha \leftarrow \textsc{Clamp}(v.\alpha, 0.0, 1.0)$\;
        
        $P \leftarrow \textsc{Lerp}(v.\text{Prev}.\text{pos}, v.\text{Target}.\text{pos}, v.\alpha)$\;
        $R \leftarrow \textsc{Lerp}(v.\text{Prev}.\text{rot}, v.\text{Target}.\text{rot}, v.\alpha)$\;
        
        \textsc{SetActorTransform}$(P, R)$\;
    }
}
\end{algorithm}

The interpolation algorithm is presented as pseudocode in \cref{alg:interpolation}. \textsc{OnNetworkUpdate} is invoked by the producer thread everytime it receives a TraCI response packet from the SUMO server, while \textsc{TickVisuals} is invoked by the consumer thread whenever it updates the visual state. Note that this algorithm works regardless if the simulation rate $N$ is greater than, equal to or smaller than the framerate.

\subsubsection{Coordinate Transformation}
While the interpolation scheme ensures temporal smoothness, correct spatial alignment between the simulation and the rendering engine is also important. Before starting the simulation, we reproject the SUMO road network into the Swiss national coordinate system (EPSG:2056) to ensure it aligns with common 3D models of Zurich. Subsequently, We convert 2D Cartesian coordinates received from the SUMO server into Unreal Engine’s 3D space by subtracting predefined offsets from the raw positions and inverting the Y-axis to account for different orientations of the coordinate system. Vehicle orientation is corrected by incorporating a 90 degrees offset, while the vertical position is dynamically determined to ensure vehicles snap accurately to the underlying terrain. 

\subsubsection{Raycasting for Height Finding}
\label{sec:raycast}
To vertically align vehicles with the 3D terrain, the vertical coordinate is derived dynamically via a top-down raycast. For each vehicle, a ray is cast from a fixed height to intersect with relevant environmental geometry, such as the city's basemap mesh, allowing the actor to snap to the road surface.

Given that the target terrain features only very gradual elevation changes, a single raycast at the vehicle's center proved sufficient, but the approach could be adapted to steeper terrain by performing a dual raycast, one at the front and one at the rear of the vehicle, to calculate an appropriate pitch rotation. 
% During initial testing, we encountered discrepancies between the visual geometry and the calculated collisions, causing vehicles to float above or clip into the ground. This problem could be traced to a simplified proxy mesh generated by Unreal Engine's Nanite system, which was used for collision detection. Disabling nanites for the affected meshes resolves the issue.

\subsection{Traffic Light Data Processing}
In contrast to vehicles, which can dynamically enter and exit the simulation, traffic lights are persistent infrastructure elements. Consequently, their physical representation is instantiated during the initialization phase, and only their signal states are updated dynamically. SUMO defines traffic lights as part of an abstract road network definition in the form of an XML file. Our framework parses this definition to instantiate simplified traffic light meshes and establish the necessary mapping between the simulation and the 3D actors.

\subsubsection{Network Parsing and Instantiation}
To instantiate traffic lights correctly, we first need to understand how they are represented in the SUMO network definition and how they are controlled during the simulation. The network is defined by a root \textit{net} node with all relevant entities being direct children of it. Traffic lights are clustered as a type of \textit{junction} and are associated with a signal schedule, represented as strings of the form 'GrGr', where the ith character represents the signal for the ith traffic light of the corresponding junction. Each \textit{junction} has incoming and outgoing \textit{lanes}. \textit{Lanes} are connected via \textit{connections}, which have the \textit{tl} and \textit{linkIndex} attributes if they are part of a traffic light-controlled junction, containing the ID of that junction and the connection's index in the traffic light schedule, respectively.

Our system iterates over all children of the network root to find all connections linked to a traffic light-controlled junction as well as all lanes. For every such connection, we find its incoming lane as well as the lane's shape, defined as a list of 2D points. We spawn a traffic light actor at the shape's last point, add a default height offset, and rotate it to face the shape's second-to-last point. Finally, we add the spawned actor to a $$Map \langle JunctionID, Map \langle LinkIndex, Actor \rangle \rangle$$ mapping to support simulation-time state synchronization. The procedure is presented as pseudocode in \cref{alg:tl_init}.

\begin{algorithm}[t]
\caption{Traffic Light Initialization}
\label{alg:tl_init}
\SetKwProg{Fn}{Function}{}{}
$M_{TL} \leftarrow$Map$\langle$JunctionID, Map$\langle$LinkIndex, Actor$\rangle\rangle$;
\BlankLine
\Fn{\textsc{SpawnTrafficLights}(\text{NetworkXML})}{
    $\mathcal{C} \leftarrow \textsc{ParseTLConnections}(\text{NetworkXML})$\;
    $\mathcal{L} \leftarrow \textsc{ParseLanes}(\text{NetworkXML})$\;
    \BlankLine
    \ForEach{$c \in \mathcal{C}$}{
        $l \leftarrow \mathcal{L}[c.\text{laneIn}]$\;
        $P_{\text{last}}, P_{\text{prev}} \leftarrow \textsc{GetLastPoints}(l.\text{shape})$\;
        \BlankLine
        $Pos \leftarrow \textsc{ToUnreal}(P_{\text{last}}) + (0, 0, H_{\text{offset}})$\;
        $Rot \leftarrow \textsc{RotationFromTo}(P_{\text{prev}}, P_{\text{last}})$\;
        \BlankLine
        $A \leftarrow \textsc{SpawnActor}(Pos, Rot)$\;
        $M_{TL}[c.\text{tlID}][c.\text{linkIdx}].\textsc{Add}(A)$\;
    }
}
\end{algorithm}

\subsubsection{State Synchronization and Visualization}
As part of the producer-consumer approach outlined in \cref{sec:prodcons}, we receive junction-based traffic light data in the form of strings of the form 'GrGr' from the SUMO server. Thanks to the previous instantiation step, we can directly map this state information to the corresponding actor, where we update the light mesh's color accordingly.

\subsection{Optimizations}
Maintaining a high, stable frame rate is critical for immersion in virtual reality. Since the SUMO simulation often includes thousands of vehicles across the city network, a naive implementation that instantiates and updates an actor for every simulated entity would introduce a severe performance bottleneck. We implement a number of optimizations to mitigate this.

\subsubsection{Distance Culling and Scheduling}
In dense urban environments, distant vehicles are usually obstructed by building geometry and don't contribute to the user experience. To exploit this, we only handle vehicles within a configurable distance from the player. When processing simulation updates, we categorize vehicles based on their proximity: we spawn actors for those entering the radius, update existing ones staying within the radius, and despawn those exiting it.

Additionally, these distance calculations, as well as the height-finding raycasts (\cref{sec:raycast}), are performed at scheduled intervals rather than every frame, reducing overhead without impacting visual fidelity.

\subsubsection{Object Pooling}
To avoid the computational cost of frequently creating and destroying actors, we implement an object pool pattern. A collection of vehicle actors is instantiated in an inactive state at the start of the simulation. When a vehicle enters the culling radius, an actor is retrieved from the pool and activated; when it exits, it is deactivated and returned to the pool rather than being destroyed. The system maintains separate pools for different vehicle types and supports dynamic resizing should the traffic volume exceed the initial buffer.

\subsection{OSC Auralization}
To support the auditory component of the simulation, the framework streams relevant data out of the application via the OSC protocol. This decoupling allows the visualization and audio systems to operate asynchronously while enabling arbitrary downstream audio processing.

\subsubsection{Message Structure}
We construct and send a single aggregated OSC bundle at a fixed time interval. This bundle begins with a header containing the simulation timestamp and the player's listener position. Following the header, the system appends a list of vehicle properties for each active vehicle fulfilling one of the following conditions:
\begin{enumerate}
    \item It is being transmitted for the first time.
    \item Its position or velocity has changed beyond a configurable threshold ($\delta_{pos}, \delta_{vel}$) compared to the last transmission.
    \item A maximum "keep-alive" time interval has elapsed.
\end{enumerate}
For every such vehicle, we transmit the vehicle's ID, position, velocity, and acceleration.

\subsubsection{Audio Integration}
The resulting OSC stream provides a standardized interface for auralization. For the purpose of our user study, these messages were consumed by a custom JUCE \cite{JUCE} environment provided by the research group, which handles the synthesis and spatialization of traffic noise based on the received kinematic data.

\section{User Study}
\label{sec:experiments}

% Sell the results as:
% 1. Showing our implementation has good immersion in general 
% 2. Showing the user experiences are "perceptually aligned" with ground truth from simulation, meaning our implementation conveys the traffic situation well
% 3. Sound changes perception of safety a lot, meaning our approach of multimodality in traffic simulations is truly important

To evaluate our implementation, we conducted a user study with 20 participants. While different 3D representations of the city of Zurich are supported, we used Google's photorealistic 3D tiles \cite{googlePhotorealisticTiles} for the user study, as they cover all relevant areas of the city while being of reasonable visual quality. A junction representative of the city traffic was chosen for the study. Users were first shown a 2D visualization of traffic at the junction from the SUMO GUI. Subsequently, they were asked to put on a HTC VIVE VR headset \cite{HTCVive2016} and explore two different traffic scenarios at the junction. Scenario 1 takes place in the morning and has slower moving vehicles, while scenario 2 is in the evening with faster moving vehicles. Each scenario was first shown with sound from the auralization engine, then without sound, resulting in 4 different combinations. The participants were free to move and look around using a handheld controller as well as by turning their head physically at all times.
% A screenshot of the experiment environment is shown in \cref{?}.

After the session, participants completed a 22-item questionnaire assessing perceived safety, realism, and immersion on a 5-point Likert scale. The complete questionnaire structure is presented in \cref{tab:questionnaire}.

\begin{table*}[t] % TODO: Maybe also add the mean + std scores in the table?
    \centering
    \caption{User Study Questionnaire. Questions regarding Safety, Willingness to Cross, and Attractiveness were repeated for all 4 permutations (Morning/Evening $\times$ Sound/No Sound). All responses were recorded on a 5-point Likert scale (1=Low, 5=High), except for open-ended feedback.}
    \label{tab:questionnaire}
    \begin{tabular}{l p{10cm}}
        \toprule
        \textbf{Category} & \textbf{Question Item} \\
        \midrule
        \textbf{Comparison} & Did the 3D simulation improve the experience over viewing the 2D simulation? \\
        \midrule
        \textbf{Perceived Safety} & Overall, how safe did you feel as a pedestrian during the [first/second] scenario [with/without] sound? \\
        & \textit{(Repeated for all 4 conditions)} \\
        \midrule
        \textbf{Behavior} & How willing would you be to cross the street with a child or limited visibility in [first/second] scenario [with/without] sound? \\
        & \textit{(Repeated for all 4 conditions)} \\
        \midrule
        \textbf{Attractiveness} & How attractive did you find the [first/second] scenario [with/without] sound as a pedestrian? \\
        & \textit{(Repeated for all 4 conditions)} \\
        \midrule
        \textbf{Sound Impact} & How much did sound increase your sense of realism? \\
        & How much did sound affect your sense of safety? \\
        \midrule
        \textbf{Realism \& Fidelity} & Overall, how much did the cars in the virtual environment behave and move like they were real? \\
        & Would it be an improvement to the simulation if the cars reacted to your actions? \\
        \midrule
        \textbf{Presence} & I had a sense of ``being there'' in the traffic scenario. \\
        & To what extent were there times during the experience when the virtual traffic situation became the ``reality” for you, and you almost forgot about the ``real world” in which the whole experience was really taking place?  \\
        & When you think back about your experience, do you think of the city more as images that you saw, or more as somewhere you walked through? \\
        \midrule
        \textbf{Qualitative} & What were the most noticeable differences between the two scenarios for you? \\
        & Additional comments or suggestions. \\
        \bottomrule
    \end{tabular}
\end{table*}

\subsection{Quantitative Results} 

\subsubsection{Immersion}
The transition from the 2D SUMO GUI to the 3D immersive environment was received positively, with participants reporting a substantial improvement in the experience ($M=4.71, SD=0.47$). General immersion scores were medium-high. Participants reported a strong sense of "being there" ($M=3.85, SD=1.18$) and felt the virtual cars behaved realistically ($M=3.65, SD=0.88$). Participants also agreed that having the vehicles react to the player position would improve the experience ($M = 4.40, STD=0.88$), providing a potential future direction to explore.

\subsubsection{Comparison across scenarios}
\begin{figure*}[t]
    \centering
    \includegraphics[width=1\linewidth]{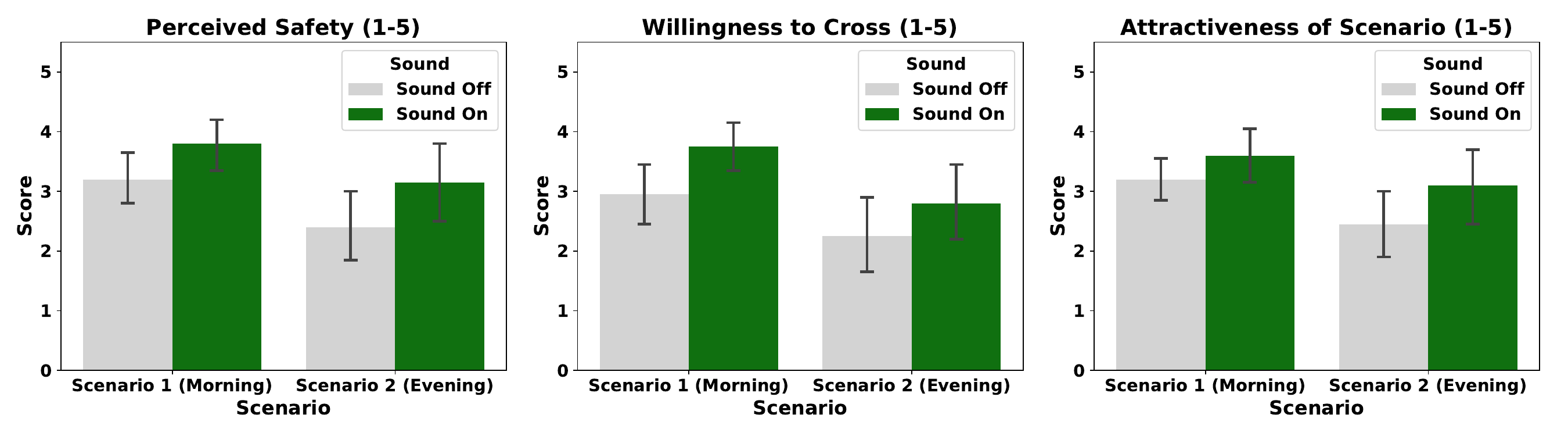}
    \caption{User study results comparing the Morning (slow traffic) and Evening (fast traffic) scenarios. The charts display mean participant ratings and error bars represent 95\% confidence intervals. The results demonstrate that the addition of spatial audio consistently improved user perception of safety, willingness to cross the road and general attractiveness of the junction for pedestrians. Furthermore, the significant drop in safety ratings for the Evening scenario confirms that participants successfully perceived the increased traffic risk in the virtual environment.}
    \label{fig:scenario_compare}
\end{figure*}

To validate the simulation's ability to convey different traffic conditions, we compare user responses between the morning (Scenario 1, slow traffic) and evening (Scenario 2, fast traffic) scenarios for the relevant questions. The results are presented in \cref{fig:scenario_compare}. Furthermore, we conducted a paired t-test, confirming that participants felt significantly less safe in Scenario 2 compared to Scenario 1, regardless if the sound is off ($M_{S2}=2.40$ vs $M_{S1}=3.20$, $p = 0.025$) or on ($M_{S2}=3.15$ vs $M_{S1}=3.80$, $p = 0.015$). The clear perceptual alignment with the dynamics of the traffic simulation demonstrates that the visual and auditory signals successfully communicated the varying levels of danger.

\subsubsection{Impact of Audio}
The addition of spatial audio had a measurable impact on user behavior and perception. Participants reported that sound significantly increased their sense of realism ($M=4.00, SD=1.34$). Notably, the presence of sound increased the perceived safety in general ($M=3.95, STD=1.10$). The feedback suggests that audio provided critical spatial awareness of out-of-view vehicles, allowing users to make more confident crossing decisions.

\subsection{Qualitative Results}

Participants provided feedback on their experience in natural language for the qualitative questions. On the difference between the two scenarios, one participant noted, ``In the second scenario, I felt much more uncomfortable to move, because the vehicles were much faster". Others mentioned that they ``could get a feel for how intense (or not) traffic was". This suggests that the combination of visual speed and spatial audio provided an intuitive understanding of the traffic volume.

Some participants noted limitations related to mesh resolution and tile transitions regarding the photorealistic 3D tiles from Google. A few cases of dizziness while wearing the HTC Vive headset were also reported. However, the system was overall described as stable and convincing for both visualization and interaction.

\subsection{Discussion}

Overall, the results suggest that our proposed framework effectively bridges the gap between abstract traffic simulation and high quality VR environments, reaching a good level of immersion. The significant difference in risk perception between scenarios indicates a high degree of perceptual alignment between user experience and the simulated traffic dynamics. Furthermore, the clear correlation between spatial audio and willingness to cross shows the necessity of multimodal feedback in traffic simulation. 

We hypothesize that the dizziness reported by some participants stems from the inherent limitations of the outdated HTC Vive hardware, specifically its low resolution and pronounced screen door effect. Another possible, but less likely explanation is that the end-to-end latency in the entire pipeline exceeded the threshold for comfortable VR at times: While the HTC Vive targets a refresh rate of $90$ Hz and a latency of $21$-$42$ ms \cite{HTCVive2016}, the SUMO server can fail to keep up with the simulation step frequency due to hardware constraints, leading to an increase in total latency.
\section{Conclusion and Future Work}
\label{sec:conclusion}
\subsection{Conclusion}
In this work, we presented a real-time visualization framework that couples microscale traffic simulations with an immersive virtual model of Zurich. By bridging the gap between SUMO and Unreal Engine 5, we achieved a high quality representation of traffic dynamics that operates independently of the game engine framerate and supports the rendering of simulated traffic lights. Furthermore, we implemented an OSC interface, allowing plug-and-play integration of auralization engines.

Our user study showed that the proposed framework provided a strong sense of immersion and achieved a high degree of perceptual alignment between simulated traffic dynamics and human judgment of safety. Notably, the inclusion of spatial audio was found to be a decisive factor in users' sense of presence and safety, indicating it as an important aspect of virtualized participatory urban planning.

\subsection{Limitations and Future Work}
While the current framework successfully visualizes traffic flow, it relies on a uni-directional coupling; the SUMO server drives the visualization, but the VR user cannot physically interact with the traffic logic (e.g., a vehicle will not brake if the user steps onto the road). Future iterations could implement a bi-directional coupling to enable real-time pedestrian-vehicle interaction. Additionally, a higher resolution 3D model of the city could be acquired, and a better VR headset could be used for user studies. 

\subsection{Strengths and Weaknesses}
We end this work by summarizing three strengths and weaknesses of our framework.

\paragraph{Three Key Strengths.}
\begin{itemize}
  \item High-fidelity coupling between microscale traffic simulation and immersive VR visualization.
  \item Decoupled, asynchronous C++ architecture ensuring temporal stability and real-time performance.
  \item Integration of spatialized audio enabling multimodal perception of traffic dynamics.
\end{itemize}

\paragraph{Three Key Weaknesses.}
\begin{itemize}
  \item Uni-directional coupling prevents real-time pedestrian-vehicle interaction.
  \item Dependence on external photorealistic city tiles limits generalization to other cities.
  \item Limited quantitative evaluation of simulator sickness.
\end{itemize}

{
    \small
    \bibliographystyle{unsrtnat}
    \bibliography{main}
}


\end{document}